\documentclass[aps,physrev,twocolumn,superscriptaddress]{revtex4-2}

\usepackage{graphicx}
\usepackage{float}
\usepackage{appendix}
\usepackage{amsmath}
\usepackage{amssymb}
\usepackage{svg}
\usepackage{xcolor}
\usepackage{bookmark}
\usepackage{qcircuit}
\usepackage{acronym}

\usepackage{changes} 

\begin{document}



\title{Simulating the Dynamics of Markovian Quantum Processes by Quantum Collision Models on Quantum Computers}


\author{Zeqing Wang}
\email{zeqing.wang@riken.jp}
\affiliation{Computational Materials Science Research Team, RIKEN Center for Computational Science (R-CCS), Kobe, Hyogo 650-0047, Japan}

\author{Julian D. Teske}
\email{julian.teske@q-ctrl.com}

\affiliation{%
 Q-CTRL, Berlin, Germany
}

\author{Anshuman Bhardwaj}

\affiliation{Computational Materials Science Research Team, RIKEN Center for Computational Science (R-CCS), Kobe, Hyogo 650-0047, Japan}

\author{Masahiro O. Takahashi}

\affiliation{Computational Materials Science Research Team, RIKEN Center for Computational Science (R-CCS), Kobe, Hyogo 650-0047, Japan}

\author{Seiji Yunoki}

\affiliation{Computational Materials Science Research Team, RIKEN Center for Computational Science (R-CCS), Kobe, Hyogo 650-0047, Japan}
\affiliation{%
Quantum Computational Science Research Team, RIKEN Center for Quantum Computing (RQC), Wako, Saitama 351-0198, Japan
}%
\affiliation{%
Computational Condensed Matter Physics Laboratory, RIKEN Pioneering Research Institute (PRI), Wako, Saitama 351-0198, Japan
}%
\affiliation{%
Computational Quantum Matter Research Team, RIKEN Center for Emergent Matter Science (CEMS), Wako, Saitama 351-0198, Japan
}%


\date{\today}

\begin{abstract}
Hamiltonian dynamics have been widely implemented on 
noisy intermediate-scale quantum devices in recent years.
In contrast, experimental demonstrations of Markovian quantum dynamics remain limited, because implementing nonunitary evolution on quantum computers is challenging. 
Quantum collision models provide a natural approach to this problem by coupling the system to ancillas to realize dissipation. 
However, previous implementations of quantum collision models on quantum computers have typically been restricted to one or two system qubits and fewer than 12 time steps, owing to noise, circuit depth, the overhead of ancilla reset, and limited qubit resources. 
In this work, we experimentally simulate Markovian quantum processes with local and nonlocal dissipation on both trapped-ion and superconducting quantum computers. 
By employing hardware-specific ancilla strategies, we realize simulations with up to seven system qubits, corresponding to 13 qubits in total, and 40 time steps. 
Our results demonstrate that, even for the same physical model, the optimal implementation strategy depends strongly on the hardware characteristics of the quantum computer.
\end{abstract}


\maketitle


\section{Introduction}
Quantum computing has become an increasingly active field in recent years due to the rapid progress in experiments on different platforms,
such as superconducting~\cite{Arute2019, Kim2023},
trapped ion~\cite{hughes2025trappediontwoqubitgates9999, ransford2025helios98qubittrappedionquantum},
neutral atom~\cite{s41586-025-09848-5, s41586-025-09596-6, s41586-025-09475-0} and photonic systems~\cite{s41586-024-08406-9}.
However, although the qubit number has already reached more than one hundred with steadily declining error rates, significant progress is still required before a full \ac{FTQC} is available~\cite{2024-nature-ibm, 2025-nature-google}.
Simultaneously, the classical simulation algorithms for quantum hardware have made great progress~\cite{10.21468/SciPostPhys.15.6.222, PRXQuantum.5.010308,  doi:10.1126/sciadv.adk4321}, which makes quantum advantage a moving target~\cite{schirber2024moving}.

While \ac{FTQC} remains out of reach,
\ac{NISQ} devices have already been explored across a variety of fields facilitated by \ac{QES}~\cite{Benchmarking_Fire, QES_tomog_White2020, QES_tomog_PRXQuantum.3.020344, QES_tomog_White2025whatcanunitary, QES_dd_1998, QES_dd_survey} and \ac{QEM} methods~\cite{Cai2021, doi:10.7566/JPSJ.90.032001, PhysRevE.104.035309, RevModPhys.94.015004, RevModPhys.95.045005, fire_opal}.
Many quantum phenomena have been experimentally observed on \ac{NISQ} devices such as time crystals~\cite{PRXQuantum.2.030346, Mi2022, doi:10.1126/sciadv.abm7652, Shinjo2026},
topological phases~\cite{doi:10.1126/science.adp6802},
superconducting pairing correlations~\cite{granet2025superconductingpairingcorrelationstrappedion}
and edge modes~\cite{doi:10.1126/science.abq5769}.
Beyond many-body physics, NISQ devices have also found applications in quantum chemistry~\cite{RevModPhys.92.015003}, including studies of strongly correlated molecules~\cite{doi:10.1126/sciadv.adu9991,Shirakawa2025} and biomolecules comprising increasingly large numbers of atoms~\cite{Merz2026}.

An open quantum system~\cite{breuer_theory_2009} is a system that interacts with an external environment, exchanging energy and information. 
After tracing out the environment, the effective description of the system generally includes dissipation, which typically drives the system toward a mixed state.
Therefore, describing the states of an open quantum system requires density matrices rather than merely wave functions.
Since the size of density matrices scales quadratically with the Hilbert space dimension of the system, in contrast to the linear scaling of wave function~\cite{2024-PRXQuantum-tutorial}, simulating the dynamics of an open quantum system usually is more costly on a classical computer.
Consequently, it is natural to simulate open quantum systems on a quantum computer.
However, implementing such simulations on quantum computers is also challenging, because open quantum system dynamics are inherently non-unitary, whereas only unitary gates can be directly realized on a quantum computer.

There are several methods to realize the non-unitary operations on quantum computers~\cite{doi:10.1021/acs.chemrev.4c00428},
which can be broadly divided into two categories depending on whether the Hilbert space for the non-unitary operator is dilated by ancillary qubits.
For example, quantum imaginary time evolution~\cite{s41567-019-0704-4, 2022-PRX-qite} and variational algorithms~\cite{2019-npj-variational-ite, Watad_2024} approximate the non-unitary step by a state-dependent unitary operation without dilating the Hilbert space for the non-unitary operator.
Another category is by block-encoding, or dilation process to map a non-unitary evolution into a unitary evolution in a larger Hilbert space, performed by using additional ancillary qubits~\cite{s41467-025-55953-4, doi:10.1126/science.adh9932}.
For example, an open quantum system comprising 86 system qubits was recently simulated on a quantum computer with up to three Trotter steps, by specially designing ancilla-assisted dissipative channels for the single-qubit and two-qubit dissipators in their model~\cite{tirado2026utilityscalequantumexperimentsusing}.

The quantum collision model~\cite{PhysRevA.96.032107, CICCARELLO20221}, also known as repeated interaction scheme, is a physically intuitive description for open quantum systems, which is theoretically equivalent to the conventional system-environment scenario.
The quantum collision model describes a large environment through repeated interactions with many small, identical auxiliary systems.
Experimentally, quantum collision models have been realized in trapped-ion simulators~\cite{nature09801, nphys2630} and optical systems~\cite{s41598-019-39832-9}.
Regarding implementations on quantum computers, several theoretical proposals have been made in recent years~\cite{PhysRevA.83.062317, PhysRevLett.126.130403, 3trk-smbh, pocrnic2025quantumsimulationlindbladiandynamics}.
However, the realization on real hardware is limited to cases with one or two system qubits and fewer than 12 time steps~\cite{García-Pérez2020, PRXQuantum.4.010324, Li_2026}.

In this work, we experimentally realize dynamical simulations of Markovian quantum processes governed by Lindblad master equations on one trapped-ion quantum computer and two superconducting quantum computers, covering both local and nonlocal dissipation. 
With the aid of \ac{QEM} and \ac{QES}, we implement systems with up to 13 qubits, consisting of 7 system qubits and 6 ancillas, and simulate dynamics for up to 40 time steps. 
By adopting hardware-specific ancilla strategies, both types of quantum computers yield results in good agreement with numerical results of the corresponding Lindblad master equations.

The rest of this paper is organized as follows.
We first provide a brief review of the quantum collision models for the Markovian quantum processes and its implementation on quantum computers in Sec.~\ref{sec:qcm-qc}.
In Sec.~\ref{sec:tls}, we demonstrate the quantum collision model for the simple yet important spontaneous emission of a two-level system.
In Sec.~\ref{sec:nonlocal}, we experimentally realize the quantum \ac{ASEP} which has nonlocal dissipation.
Finally, we summarize this work and discuss future applications.
In Appendix~\ref{sec:proof}, we derive the Lindblad master equation from quantum collision models.
In Appendix~\ref{sec:estimators}, we describe the estimators used in this work. 
In Appendix~\ref{sec:hybrid_ancilla} and Appendix~\ref{sec:circuit-detail}, we provide details of our strategies and circuit constructions. 
Supplementary experimental data are presented in Appendix~\ref{sec:further-exp-data}. 
Finally, a list of acronyms used in this paper is provided in Appendix~\ref{sec:acronyms}.

\section{Quantum-collision-model framework\label{sec:qcm-qc}}

\subsection{Quantum collision models for Markovian quantum processes on quantum computers}

In many experimentally relevant settings, open quantum system dynamics can be well approximated as Markovian, i.e., as an effectively memoryless process.
In this regime, the dynamics is described by the \ac{GKSL}, or Lindblad, master equation on a coarse-grained timescale $\Delta t$~\cite{cohen-tannoudji_atom-photon_1992},
\begin{align} \label{eq:lindblad-eq}
    \frac{\hat{\rho}(t + \Delta t) - \hat{\rho}(t)}{\Delta t}
    = \mathcal{L}[\hat{\rho}(t)],
\end{align}
where $\hat{\rho}(t)$ is the density operator of the system at time $t$. 
Here, $\mathcal{L}$ is the Lindblad generator of the irreversible dynamical semigroup governing the time evolution of the system (we set $\hbar = 1$): 
\begin{align}\nonumber
    \mathcal{L}[\hat{\rho}(t)]
    =& -i [\hat{H}, \hat{\rho}(t)] \\ \label{eq:lindblad}
    &+ \sum_j \Gamma_j
    \left[
        \hat{L}_j \hat{\rho}(t) \hat{L}_j^{\dagger}
        - \frac{1}{2} \left\{ \hat{L}_j^{\dagger} \hat{L}_j, \hat{\rho}(t) \right\}
     \right],
\end{align}
where $\{\hat{A}, \hat{B}\}\equiv \hat{A}\hat{B} + \hat{B}\hat{A}$ denotes the anticommutator.
The coarse-grained timescale $\Delta t$ should be much longer than the environment correlation time $\tau_c$, so that the environment can be regarded as effectively reset to the same state at each coarse-grained time step.
At the same time, $\Delta t$ should be much shorter than the characteristic system-evolution timescale $T_R$, so that the system dynamics remains smooth on this timescale. 
Therefore, the coarse-grained timescale $\Delta t$ must satisfy,
\begin{align}
    \tau_c\ll \Delta t \ll T_R.
\end{align}

The basic idea of quantum collision models is to represent a large environment by a sequence of small, identical ancillas that interact with the system successively, with each ancilla interacting with the system over a coarse-grained time interval $\Delta t$.
From the perspective of the system, these repeated interactions are effectively indistinguishable from interacting with a large environment that rapidly loses memory of the interaction and returns to the same state on the coarse-grained timescale $\Delta t$. 
Consequently, after tracing out either the ancillas or the large environment, the reduced system undergoes the same effective dynamics, which can be described by a Lindblad master equation. 
The equivalence between these two microscopic descriptions of open quantum systems has been established theoretically~\cite{CICCARELLO20221}.

As proposed in Refs.~\cite{PhysRevLett.126.130403, 3trk-smbh}, quantum collision models can be used to efficiently simulate Markovian quantum dynamics on quantum computers.
The time evolution of an open quantum system governed by the Lindblad equation, Eq.~\eqref{eq:lindblad-eq}, can be decomposed into two parts.
The first is the unitary evolution governed by the system Hamiltonian $\hat{H}$ in Eq.~\eqref{eq:lindblad}.
This part can be implemented by Trotterizing the time-evolution operator $e^{-i\hat{H} t}$ into small time steps, a standard approach that has already been realized for a variety of Hamiltonians. 
The second is the non-unitary evolution induced by the environment, which is described by the jump operators $\hat{L}_j$ in Eq.~\eqref{eq:lindblad} with rates $\Gamma_j$. 
Quantum collision models implement this non-unitary part by introducing ancillas and evolving the system and ancillas jointly.

\begin{figure*}
\includegraphics[width=.8\linewidth]{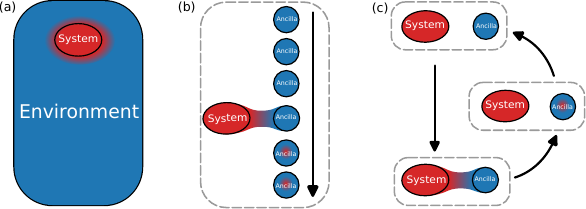}%
\caption{
    (a) Schematic of the conventional system-environment description of an open quantum system.
    (b,c) Schematics of two ancilla strategies for quantum collision models with a single jump operator.
    (b) Fresh-ancilla strategy, in which a fresh ancilla is used after each system-ancilla interaction.
    (c) Reset-ancilla strategy, in which the same ancilla is reused after being reset.
    \label{fig:collision-model}}
\end{figure*}

In the conventional system-environment description, a jump operator $\hat{L}_j$ is a system operator that arises from decomposing the system-environment interaction into components associated with the Bohr frequencies of the system. 
It connects energy eigenstates whose Bohr-frequency differences are resonant or near-resonant with frequencies supported by the environment.
The rapid loss of memory in the environment makes the reduced dynamics of the system irreversible after the environment is traced out. 
As an example, consider the spontaneous emission from a two-level system induced by a dipole interaction with the vacuum electromagnetic field. 
The system operator in the dipole interaction contains both $\hat{\sigma}^+ = |1\rangle \langle 0|$ and $\hat{\sigma}^- = |0\rangle \langle 1|$, which raise and lower the two-level system between the ground state $|0\rangle$ and the excited state $|1\rangle$, respectively. 
Since the electromagnetic field is infinitely large and initially in the vacuum state, any photon emitted by the two-level system rapidly escapes into the environment. 
As a result, the two-level system effectively continues to interact with a vacuum environment, and the absorption process associated with $\hat{\sigma}^+$ is negligible.
The corresponding jump operator is therefore $\hat{L} = \hat{\sigma}^-$.

At this point, it is natural to use a qubit initialized in the ground state $|0\rangle$ as the ancilla associated with the jump operator $\hat{L}_j$, and to choose the system-ancilla interaction as 
\begin{align}\label{eq:interaction}
    \hat{V}_j = g_j(\hat{L}_j\hat{\sigma}^+_{\mathrm{a}, j}
     + \hat{L}_j^{\dagger}\hat{\sigma}^-_{\mathrm{a}, j}),
\end{align}
where the ground and excited states of the ancilla qubit form the computational basis $\{|0\rangle, |1\rangle\}$.
The subscript $\mathrm{a}, j$ indicates that the operator acts on the $j$th ancilla, and $g_j$ denotes the coupling strength. 
For a sufficiently short time step $\Delta t \ll 1/g_j$, the ancilla remains close to the ground state $|0\rangle$. Thus, the second term in Eq.~\eqref{eq:interaction} is suppressed when the ancilla is in $|0\rangle$, since $\hat{\sigma}^-_{\mathrm{a}, j} |0\rangle = 0$. 
If the ancilla is reset to $|0\rangle$ or replaced by a fresh ancilla prepared in $|0\rangle$ after each time interval $\Delta t$, the system effectively interacts with an ancilla that is always initialized in the ground state (see Fig.~\ref{fig:collision-model}). 
This reproduces the memoryless condition of the conventional system-environment description.

In the conventional system-environment description, the dissipation rate $\Gamma_j$ is determined by the Fourier transform of the autocorrelation function of the environment operator appearing in the system-environment interaction. 
Typically, this autocorrelation function decays rapidly for time separations longer than the environment correlation time $\tau_c$. 
As a result, $\Gamma_j$ scales with the correlation time $\tau_c$ and with the square of the system-environment coupling strength.
In quantum collision models, the time step $\Delta t$ plays the role of an effective correlation time, because each ancilla interacts with the system for a duration $\Delta t$. 
This suggests that the dissipation rate in the collision models scales as $g_j^2 \Delta t$. 
Indeed, one can show that this relation holds exactly: 
\begin{align}
  \Gamma_j = g_j^2 \Delta t.  
\end{align}

Following the above analysis, the Markovian quantum dynamics governed by the Lindblad master equation, Eq.~\eqref{eq:lindblad-eq}, can be simulated by the following procedure.
At each coarse-grained time interval $\Delta t$, an ancilla initialized in the ground state $|0\rangle$ is assigned to each jump operator $\hat{L}_j$. 
The system qubits and ancillas are then evolved under 
$U(\Delta t) = e^{-i \hat{H}\Delta t}e^{-i \hat{V}\Delta t}$, where $\hat{V} =\sum_j \hat{V}_j$ and each $\hat{V}_j$ is given by Eq.~\eqref{eq:interaction}.
After the evolution over one time step $\Delta t$, the system density operator is updated by a quantum collision map $\mathcal{E}$ acting on the pre-step state,
\begin{align}\label{eq:quantum-collision-map}
    \hat{\rho}(t+\Delta t) =
    \mathcal{E}[\hat{\rho}(t)],
\end{align}
where 
\begin{align}\label{eq:collision-map}
    \mathcal{E}[\hat{\rho}(t)]
    \equiv \mathrm{Tr}_{\mathrm{a}}\left[
    U(\Delta t)
    \left(
    \hat{\rho}(t)\bigotimes_j |0_{\mathrm{a}, j}\rangle\langle 0_{\mathrm{a}, j}|
    \right)
    U^{\dagger}(\Delta t)
     \right]
\end{align}
and $\mathrm{Tr}_{\mathrm{a}}$ denotes the partial trace over all ancillas.
Provided that the collision-model parameters $g_j$ and $\Delta t$ satisfy 
\begin{align}\label{eq:conditions}
   \mathcal{O}(\hat{H})\ll g_j \ll \frac{1}{\Delta t},
   \quad g_j^2\Delta t = \Gamma_j,
\end{align}
where $\mathcal{O}(\hat{H})$ denotes the characteristic energy scale of the Hamiltonian $\hat{H}$,
the quantum collision map in Eq.~\eqref{eq:collision-map} can be expanded for small $\Delta t$ as 
\begin{align}\label{eq:collision-map-series}
    \frac{\mathcal{E}[\hat{\rho}(t)] - \hat{\rho}(t)}{\Delta t}
    = \mathcal{L}[\hat{\rho}(t)] + \mathcal{O}\left(\Delta t\right).
\end{align}
To leading order, the right-hand side of Eq.~\eqref{eq:collision-map-series} reduces to the Lindblad generator $\mathcal{L}$ defined in Eq.~\eqref{eq:lindblad}.
It is worth noting that the expansion in Eq.~\eqref{eq:collision-map-series} requires $g_j$ to scale as $1/\sqrt{\Delta t}$ so that the dissipation rates $\Gamma_j$ remain finite.
A derivation for the more general case of a non-diagonal master equation was given in Ref.~\cite{PhysRevLett.126.130403}.
In Appendix~\ref{sec:proof}, we provide the derivation for the special case relevant to this work.

\subsection{Reset-ancilla and fresh-ancilla strategies}

In experiments, we evolve the system from an initial state $\hat{\rho}(0)$ for $1$, $2$, $\cdots$, $N$ time steps, thereby obtaining the system dynamics from $t = 0$ to $t = N\Delta t$.
A key issue is how to prepare the ancillas in the ground state $|0\rangle$ at each time step, because after the system-ancilla interactions the ancillas generally no longer remain in $|0\rangle$. 
In practice, there are two strategies, as illustrated in Figs.~\ref{fig:collision-model}(b) and \ref{fig:collision-model}(c) for the case of a single jump operator. 
One strategy is to use a fresh ancilla for a given jump operator at each time step, as shown in Fig.~\ref{fig:collision-model}(b). 
The other is to reset the same ancilla after each system-ancilla interaction, as shown in Fig.~\ref{fig:collision-model}(c). 
The choice between these strategies depends on the hardware characteristics of the quantum platform.

In our experiments, we used three quantum computers across two different platforms:
IBM's superconducting quantum computers \emph{ibm\_kobe} (Heron r2) and \emph{ibm\_boston} (Heron r3), each with $156$ qubits,
and Quantinuum's trapped-ion quantum computer \emph{Reimei} with $20$ qubits.
For the superconducting quantum computers \emph{ibm\_kobe} and \emph{ibm\_boston},
a reset operation is implemented by a mid-circuit measurement followed by a conditional flip.
Because mid-circuit measurements can introduce substantial noise, frequent resets are challenging on these devices. 
Thus the fresh-ancilla strategy is preferable on this superconducting platform,
even though additional SWAP gates are required to route fresh ancillas to neighboring system qubits because of the limited connectivity. 
The choice of ancilla strategy for IBM superconducting quantum computers was also discussed in Ref.~\cite{PRXQuantum.4.010324}. 
On the trapped-ion quantum computer \emph{Reimei}, the number of available qubits is smaller, but reset operations are less noisy; therefore, the reset-ancilla strategy is preferable.
Moreover, these two strategies can be combined, as discussed in Appendix~\ref{sec:hybrid_ancilla}.
This leads to a trade-off among reset overhead, SWAP overhead, and the availability of additional ancillas.


%

\section{Simulation of local dissipation: spontaneous emission in a two-level system\label{sec:tls}}

The two-level system is one of the simplest yet most important models in physics.
Spontaneous emission occurs when a two-level system is coupled to a bosonic field in the vacuum state through an electric-dipole interaction.
As a first experiment, we simulate the spontaneous emission dynamics of a two-level system.

The Hamiltonian of the two-level system is given by 
\begin{align}
    \hat{H} = \frac{\omega}{2} \hat{\sigma}^z,
\end{align}
which can be encoded into a single qubit with
$\hat{\sigma}^z = |1\rangle\langle 1| - |0\rangle \langle 0|$.
Here, $\omega$ denotes the transition frequency.
After tracing out the bosonic field, the dynamics of this two-level system is described by the Lindblad master equation, Eq.~\eqref{eq:lindblad-eq}, with a single jump operator $\hat{L} = \hat{\sigma}^-$.
Therefore, only one ancilla is required to simulate the Markovian quantum dynamics for this two-level system.
On a quantum computer, we use one qubit to encode the two-level system and another qubit as the ancilla. 
The system-ancilla interaction is given by the spin-exchange coupling between these two qubits,
\begin{align}\label{eq:spontaneous-emission-interaction}
    \hat{V} = g ( \hat{\sigma}^-\hat{\sigma}^+_{\mathrm{a}}
    +\hat{\sigma}^+\hat{\sigma}^-_{\mathrm{a}}
    ),
\end{align}
where $g$ is the coupling strength.

On the quantum computers, the system and ancilla qubits are initially prepared in the ground state $|0\rangle$. 
We prepare the system qubit in an arbitrary spin-coherent state by applying an $R_Y$ rotation followed by an $R_Z$ rotation:
\begin{align}
    |\theta, \phi\rangle = e^{-i \frac{\phi}{2}\hat{\sigma}^z} e^{-i \frac{\theta}{2}\hat{\sigma}^y} |1\rangle
    = R_Z(-\phi) R_Y(\pi-\theta) |0\rangle,
\end{align}
where $R_Z(\alpha) = e^{-i(\alpha/2)Z}$,
$R_Y(\alpha) = e^{-i(\alpha/2)Y}$,
with $Z=-\hat{\sigma}^z$ and $Y = -\hat{\sigma}^y = i|1\rangle\langle 0| - i|0\rangle\langle 1|$.
It is worth noting that the physical Pauli operators $\hat{\sigma}^y$ and $\hat{\sigma}^z$ differ from the quantum-computing conventions $Y$ and $Z$ by an overall minus sign. 
This sign difference arises because the physical Pauli operators $\hat{\sigma}^y$ and $\hat{\sigma}^z$ are defined with $|0\rangle$ as the ground state and $|1\rangle$ as the excited state,
whereas the quantum-computing convention defines $Z = |0\rangle \langle 0| - |1\rangle\langle 1|$.

Next, we apply an $R_{XX}$ gate followed by an $R_{YY}$ gate to realize the system-ancilla interaction: 
\begin{align}
    e^{-i\hat{V}\Delta t} = R_{YY}(g \Delta t) R_{XX}(g\Delta t),
\end{align}
where $R_{XX}(\alpha)=e^{-i(\alpha/2)X\otimes X}$ and 
$R_{YY}(\alpha)=e^{-i(\alpha/2)Y\otimes Y}$, 
with $X = \hat{\sigma}^x = |1\rangle\langle 0| + |0\rangle\langle 1|$.
We then apply an $R_Z$ gate to simulate the unitary evolution generated by the system Hamiltonian:
\begin{align}
    e^{-i \hat{H} \Delta t} = R_Z(-\omega \Delta t).
\end{align}

\begin{figure}[htbp]
\includegraphics[width=\linewidth]{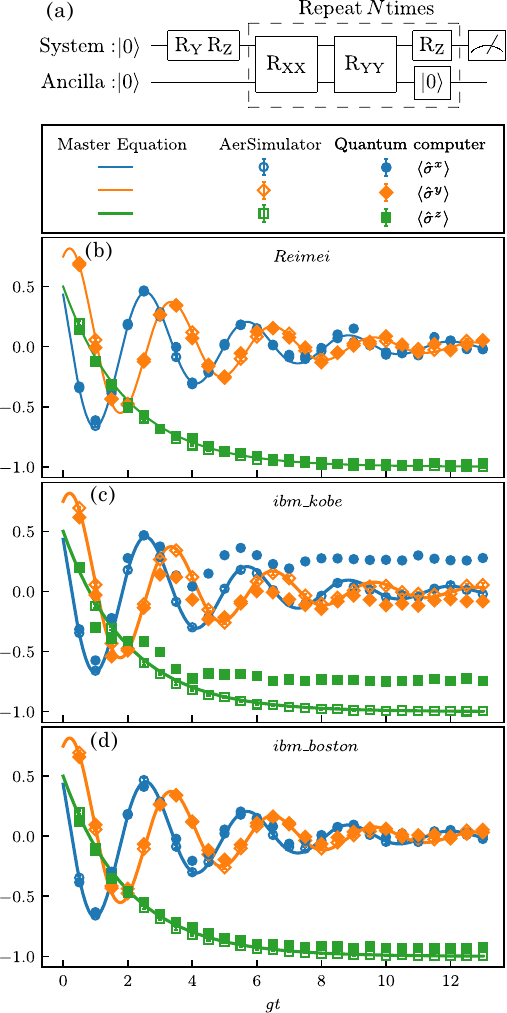}
\caption{
    (a) Circuit for simulating spontaneous emission in a two-level system using the reset-ancilla strategy. The circuit shown measures the observable $\hat{\sigma}^z$. 
    (b)--(d) Results obtained for 26 time steps on three different quantum computers: \emph{Reimei}, \emph{ibm\_kobe}, and \emph{ibm\_boston}.
    The initial state is the spin-coherent state \mbox{$\hat{\rho}(0)=|\pi/3,\pi/3\rangle\langle \pi/3,\pi/3|$}.
    Results obtained by numerically solving the Lindblad master equation and by running the circuit on the noiseless AerSimulator~\cite{qiskit2024} are also shown for reference in (b)--(d).
    The other parameters are time-step parameter $g\Delta t = 0.5$, 
    the two-level-system transition frequency $\omega = 2g$, and the number of shots $N_{\mathrm{shots}}=1024$ for \emph{Reimei} and $N_{\mathrm{shots}}=4096$ for \emph{ibm\_kobe} and \emph{ibm\_boston}. 
    Panels (b)--(d) share a common x-axis.
    \label{fig:2bits}}
\end{figure}

For this experiment, 
we use the reset-ancilla strategy on all quantum computers, in which the ancilla is reset to the ground state $|0\rangle$ after each system-ancilla spin-exchange interaction, as illustrated in Fig.~\ref{fig:collision-model}(c). 
To obtain the result at time $t = N\Delta t$, we run a circuit that repeats the system-ancilla interaction, the system Hamiltonian evolution, and the ancilla reset for $N$ time steps, as shown in Fig.~\ref{fig:2bits}(a).
The reset operation in the final time step is omitted because it is unnecessary.

At the end of the circuit, we measure three observables of the system qubit: $\hat{\sigma}^x$, $\hat{\sigma}^y$, and $\hat{\sigma}^z$.
To measure $\hat{\sigma}^z$, we perform a measurement in the computational basis, as shown in Fig.~\ref{fig:2bits}(a).
To measure $\hat{\sigma}^x$, we rotate the measurement basis from the $X$ basis to the $Z$ basis by applying a Hadamard gate, 
$H = (|0 \rangle\langle 0| + |0 \rangle\langle 1| + |1 \rangle\langle 0| - |1 \rangle\langle 1|)/\sqrt{2}$, before the computational-basis measurement. 
To measure $\hat{\sigma}^y$, we apply an S-adjoint gate,
$S^{\dagger} = |0 \rangle\langle 0| -  i |1 \rangle\langle 1|$,
followed by a Hadamard gate before the computational-basis measurement.

We now specify the parameters used in the experiment. 
We choose the time-step parameter $g\Delta t= 0.5$ and 
the transition frequency $\omega = 2 g$.
This choice corresponds to a dissipation rate $\Gamma = g^2 \Delta t = 0.5g$. 
With these parameters, both the spontaneous-emission dynamics of the population $\langle \hat{\sigma}^z\rangle$, which occurs on the timescale $\sim 1/\Gamma$,
and the oscillations of the coherences $\langle \hat{\sigma}^x\rangle$ and $\langle \hat{\sigma}^y\rangle$, which occur on the timescale $\sim 1/\omega\sim\mathcal{O}(\hat{H})$, can be observed over comparable timescales, as shown in Figs.~\ref{fig:2bits}(b)--\ref{fig:2bits}(d). 
It is worth noting that the condition in Eq.~\eqref{eq:conditions} is not satisfied for this parameter choice. 
Nevertheless, for the number of time steps considered in this experiment, the error introduced by the finite time step $\Delta t$ does not accumulate significantly, and the quantum collision model remains valid.
As shown in Figs.~\ref{fig:2bits}(b)--(d), this is confirmed by the close agreement between the results obtained by running the circuits on the noiseless AerSimulator~\cite{qiskit2024} and the reference results obtained by numerically solving the Lindblad master equation.

The results obtained on \emph{Reimei}, \emph{ibm\_kobe}, and \emph{ibm\_boston} are shown in Figs.~\ref{fig:2bits}(b), \ref{fig:2bits}(c), and \ref{fig:2bits}(d), respectively. 
The error bars represent shot noise and are very small for the number of shots used here; details are provided in Appendix~\ref{sec:estimators}. 
The results from \emph{Reimei} agree well with the predictions of the Lindblad master equation, as shown in Fig.~\ref{fig:2bits}(b). 
In contrast, the results from \emph{ibm\_kobe} show a significant deviation from the Lindblad master-equation results. 
For \emph{ibm\_boston}, which is a newer revision of the same processor family as \emph{ibm\_kobe} and has higher gate fidelity, the results are significantly improved, although they still do not reach the accuracy achieved on \emph{Reimei}. 
This difference can be attributed primarily to the noise associated with reset operations on superconducting quantum computers.

This two-qubit benchmark experiment provides two insights for subsequent experiments. 
First, on \ac{NISQ} devices, the number of implementable time steps is limited. 
As a result, the error introduced by using a relatively large time step does not accumulate significantly and can remain much smaller than the hardware noise. 
Second, even for this simple circuit, the noise associated with reset operations is already substantial on superconducting quantum computers. 
This indicates that the fresh-ancilla strategy, or a hybrid strategy combining reset and fresh ancillas, should be adopted for larger experiments on superconducting platforms.

\section{Simulation of nonlocal dissipation: quantum asymmetric simple exclusion process} \label{sec:nonlocal}

\subsection{Model and quantum circuits}

Our second experiment realizes the quantum \ac{ASEP}~\cite{Robertson_2021, PhysRevLett.126.240403, PhysRevResearch.6.033030}. 
The quantum \ac{ASEP} describes a one-dimensional lattice with dissipative hopping. 
Here, ``exclusion" means that each site can be occupied by at most one particle. 
The one-dimensional lattice can be mapped onto a spin-$1/2$ chain by the Jordan-Wigner transformation, 
\begin{align}\label{eq:j-w-trans}
    \hat{n}_j \equiv \hat{a}_j^{\dagger}\hat{a}_j
    = \frac{\hat{\sigma}^z_j + 1}{2},
\end{align}
where $\hat{a}_j$ and $\hat{a}_j^{\dagger}$ are the fermionic annihilation and creation operators at site $j$, respectively. 
Thus, an occupied site, $\hat{n}_j = 1$, corresponds to a qubit in the excited state $|1\rangle$, whereas an unoccupied site, $\hat{n}_j = 0$, corresponds to a qubit in the ground state $|0\rangle$. 
The quantum \ac{ASEP} can therefore be simulated by a spin-$1/2$ chain with nonlocal two-spin dissipation acting on neighboring spin pairs, where each spin-$1/2$ degree of freedom is encoded into a qubit on a quantum computer.
In this experiment, we consider an $M$-site quantum \ac{ASEP} in which dissipative hopping occurs only from left to right. 
After mapping the system onto a qubit chain with $M$ qubits, this dissipative hopping is described by $M - 1$ nonlocal jump operators acting on nearest-neighbor qubit pairs, 
\begin{align}\label{eq:ASEP-jump}
    \hat{L}_j = \hat{\sigma}_j^- \hat{\sigma}_{j+1}^+, \quad
    j=1, 2, \ldots , M-1,
\end{align}
which lowers the left qubit and raises the right qubit.

\begin{figure}[htbp]
    \includegraphics[width=\linewidth]{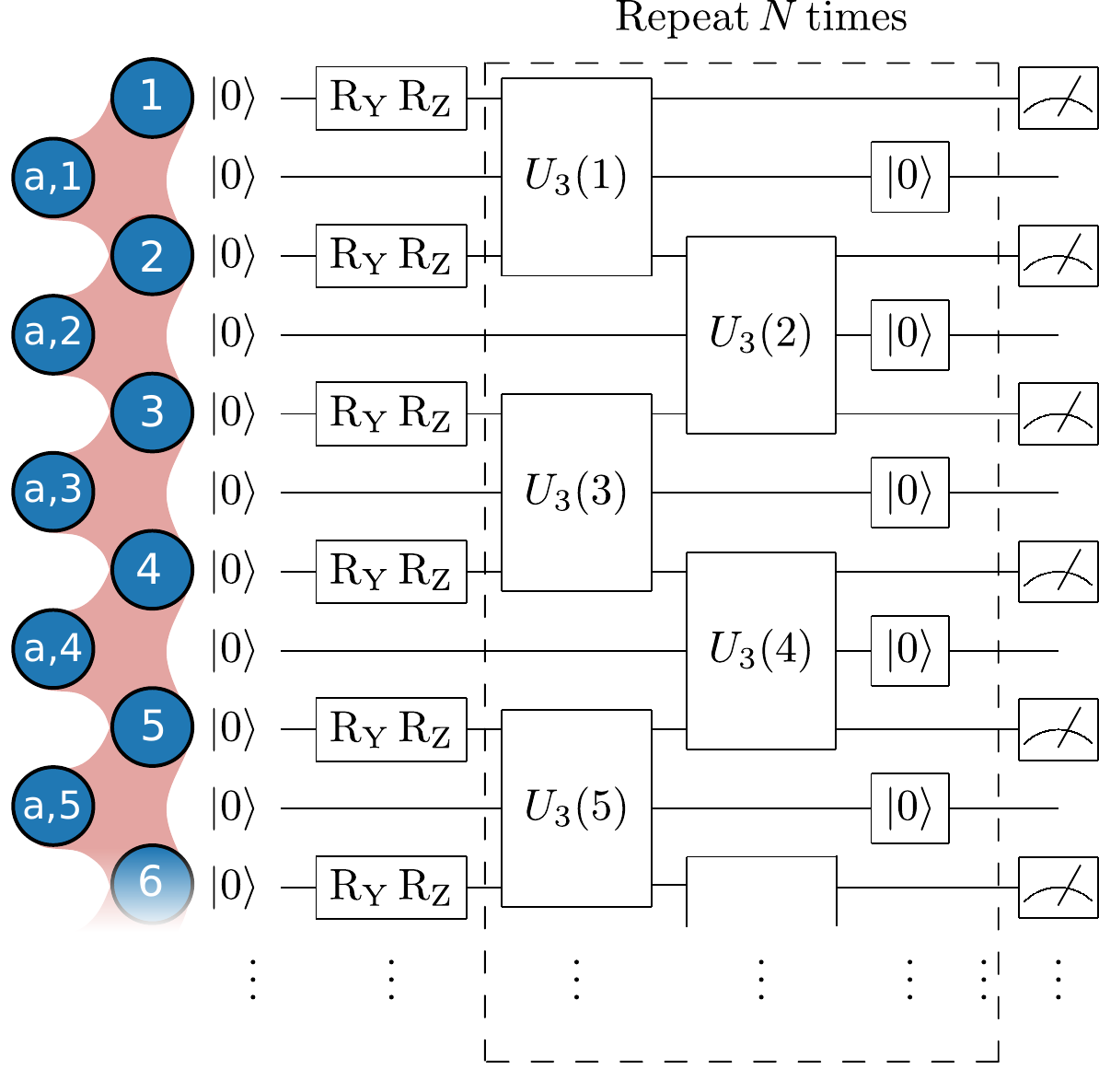}
    \caption{
        Quantum circuit for simulating quantum \ac{ASEP} dynamics. 
        The system qubits are first prepared in a product state of spin-coherent states by applying $R_Y$ and $R_Z$ gates to the initial $|0\rangle$ states.  
        The time evolution is then implemented by repeating two layers of three-qubit gates, $U_3(j)$, for $N$ time steps, corresponding to an  evolution time $N\Delta t$. 
        After each three-qubit interaction, the corresponding ancilla is reset to $|0\rangle$. 
        Finally, all system qubits are measured to obtain the expectation values $\langle\hat{\sigma}^z_j\rangle$.
        }
    \label{fig:5bits-circuit}
\end{figure}

To realize this nonlocal dissipation, we couple each neighboring spin pair to an ancilla through a three-body interaction: 
\begin{align}\label{eq:three-body-interaction}
    \hat{V}
    =\sum_{j=1}^{M-1} \hat{V}_{j}
    =\sum_{j=1}^{M-1} g_j
    \left(\hat{\sigma}_j^- \hat{\sigma}_{j+1}^+\hat{\sigma}^+_{\mathrm{a},j} + \hat{\sigma}_j^+ \hat{\sigma}_{j+1}^-\hat{\sigma}^-_{\mathrm{a},j}
    \right).
\end{align}
Thus, the circuit uses $M$ system qubits and $M-1$ ancillas, for a total of $2M - 1$ qubits. 
At the beginning of each time step, each ancilla is initialized in the ground state $|0\rangle$, so that the second term in Eq.~\eqref{eq:three-body-interaction} is suppressed. 
A schematic of this qubit chain and the corresponding quantum circuit is shown in Fig.~\ref{fig:5bits-circuit}.
Initially, all the qubits are prepared in the ground state $|0\rangle$. 
We then use $R_Y$ and $R_Z$ rotation gates to prepare the initial system state as a product state of spin-coherent states, following the same procedure as for the two-level system in Sec.~\ref{sec:tls}.

\begin{figure*}[htbp]
\includegraphics[width=\linewidth]{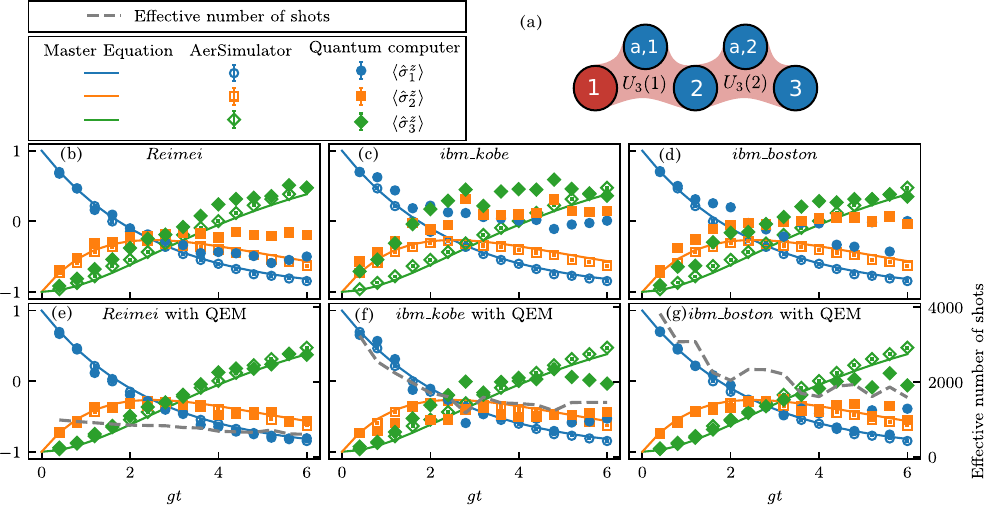}%
\caption{
    (a) Schematic of the quantum \ac{ASEP} simulation with three system qubits and two ancillas. 
    The colors of the qubits indicate the initial state: 
    blue denotes the ground state $|0\rangle$, and red denotes the excited state $|1\rangle$. 
    (b)--(g) Results obtained on three different quantum computers. 
    The system is evolved with the time-step parameter $g\Delta t = 0.4$ for $N = 15$ time steps.  
    The number of shots is $N_{\mathrm{shots}} = 1024$ for \emph{Reimei} and $N_{\mathrm{shots}} = 4096$ for \emph{ibm\_kobe} and \emph{ibm\_boston}. 
    Panels (b)--(d) show the raw results without \ac{QEM}, whereas 
    panels (e)--(g) show the results after post-selection \ac{QEM}.
    The gray dashed lines in panels (e)--(g) indicate the effective number of shots after post-selection.
    Results obtained by numerically solving the Lindblad master equation and by running the circuits on the noiseless AerSimulator are also shown for reference.
    Panels (b)--(d) share a common y-axis, 
    and panels (e)--(g) share a common y-axis. 
    Each column shares a common x-axis: (b,e), (c,f), and (d,g).
    \label{fig:5bits}}
\end{figure*}

To implement the time-evolution operator $U = e^{-i \hat{V} \Delta t}$ for the system-ancilla interaction in Eq.~\eqref{eq:three-body-interaction},
we split the interaction into an odd part,
\begin{align}
\hat{V}_{\mathrm{odd}} = \hat{V}_1 + \hat{V}_3 + \hat{V}_5 + \cdots
\end{align}
and an even part,
\begin{align}
\hat{V}_{\mathrm{even}} = \hat{V}_2 + \hat{V}_4 + \hat{V}_6 + \cdots.
\end{align}
We then use a first-order Trotter decomposition, 
$U \approx U_{\mathrm{odd}} U_{\mathrm{even}}$, 
with $U_{\mathrm{odd}}= e^{-i \hat{V}_{\mathrm{odd}} \Delta t}$ and $U_{\mathrm{even}} = e^{-i \hat{V}_{\mathrm{even}} \Delta t}$. 
Since the terms within each of $\hat{V}_{\mathrm{odd}}$ and $\hat{V}_{\mathrm{even}}$ commute with one another, $U_{\mathrm{odd}}$ and $U_{\mathrm{even}}$ can be implemented as two layers of three-qubit gates, $U_{\mathrm{odd}} = U_3(1)U_3(3)U_3(5)\cdots$
and
$U_{\mathrm{even}} = U_3(2)U_3(4)U_3(6)\cdots$,
where $U_3(j) = e^{- i \hat{V}_j \Delta t}$ acts on system qubits $j$ and $j+1$ and ancilla $\mathrm{a},j$. 
Although this layered implementation introduces an additional Trotter error, the error remains within an acceptable range, as evidenced by the close agreement between the Lindblad master-equation results and the noiseless classical circuit simulations in Figs.~\ref{fig:5bits} and \ref{fig:13bits}.
After repeating these two layers $N$ times, we measure $\hat{\sigma}^z$ on each system qubit. 
As in the spontaneous-emission experiment in Sec.~\ref{sec:tls}, reset operations are not applied in the final time step.

Decomposing the three-qubit unitaries $U_3(j)$ into sequences of native gates is a numerically challenging task~\cite{Leap3qubitSynth}, and the quality of the resulting circuits depends on the transpiler used. 
The circuits executed on \emph{Reimei} were compiled using Quantinuum Nexus~\cite{quantinuum_nexus}. The circuits executed on \emph{ibm\_kobe} and \emph{ibm\_boston} were compiled using Qiskit~\cite{qiskit2024}, except for the data shown in Fig.~\ref{fig:13bits-compare}(b), for which Q-CTRL's Fire Opal software~\cite{fire_opal} was used to provide the error suppression required for large circuits.

\subsection{Post-selection \ac{QEM}}

Since the circuit depth increases rapidly with the number of time steps $N$ owing to the costly three-qubit gates used in this experiment, 
it is critical to employ \ac{QEM} to extract reliable results. 
In general, \ac{QEM} reduces bias at the cost of increased variance~\cite{RevModPhys.95.045005}. 

For the quantum \ac{ASEP} experiment, the total $z$-magnetization of the system qubits,
\begin{align}
    \hat{S}^z =\sum_{j=1}^{M} \hat{\sigma}_j^z,
\end{align}
is conserved under the Lindblad dynamics because it lies in the kernel of the adjoint Lindblad generator $\mathcal{L}^{\dagger}$, i.e., $\mathcal{L}^{\dagger}[\hat{S}^z] = 0$, 
where $\mathcal{L}^{\dagger}$ is defined as~\cite{breuer_theory_2009},
\begin{align}
    \mathcal{L}^{\dagger} [\hat{S}^z]
    =i [\hat{H}, \hat{S}^z]
    + \sum_j \Gamma_j
    \left[
        \hat{L}_j^{\dagger}\hat{S}^z \hat{L}_j
        - \frac{1}{2} \left\{ \hat{L}_j^{\dagger} \hat{L}_j, \hat{S}^z \right\}
     \right].
\end{align}
As a result, the expectation value of the total $z$-magnetization at any time $t$,
\begin{align}
    \langle \hat{S}^z(t) \rangle = \mathrm{Tr}\left[ \hat{S}^z \hat{\rho}(t)\right],
\end{align}
is equal to its initial value, 
\begin{align}
    \langle \hat{S}^z(t) \rangle  = \langle \hat{S}^z (0) \rangle  \equiv \langle \hat{S}^z \rangle .
\end{align}

In addition, the quantum \ac{ASEP} considered here satisfies $\hat{H} = 0$ and $[\hat{S}^z, \hat{L}_j] = 0$ for all $j$, which is a stronger condition than $\mathcal{L}^{\dagger}[\hat{S}^z] = 0$.
This ensures that if the initial state is an eigenstate of $\hat{S}^z$, an ideal single-shot measurement at any time gives the same $\hat{S}^z$ eigenvalue, which is equal to the conserved value $\langle \hat{S}^z\rangle$.
Intuitively, this is because the jump operator in Eq.~\eqref{eq:ASEP-jump} describes dissipative spin exchange between two neighboring spins and therefore does not change the total $z$-magnetization. 

This enables a post-selection \ac{QEM} scheme. 
Specifically, since the initial states used in our experiments are eigenstates of $\hat{S}^z$, after the final measurement of all system qubits, we discard measurement outcomes for which the measured value of $\hat{S}^z$ differs from the conserved value. 
Although this post-selection process causes the effective number of shots to decrease rapidly, as shown by the gray dashed lines in Figs.~\ref{fig:5bits}(e)--\ref{fig:5bits}(g), Fig.~\ref{fig:13bits}(e) and Fig.~\ref{fig:13bits-compare}, this \ac{QEM} scheme is highly effective for this model.

\subsection{Experimental results for $M=3$}

We first run circuits for three system qubits on \emph{Reimei}, \emph{ibm\_kobe}, and \emph{ibm\_boston} using the reset-ancilla strategy, as shown in Figs.~\ref{fig:5bits}(b,e), \ref{fig:5bits}(c,f), and \ref{fig:5bits}(d,g), respectively. 
For this case, we evolve the system with time-step parameter $g\Delta t = 0.4$ for $N=15$ time steps. 
The initial state is chosen such that the first system qubit is in the excited state $|1\rangle$, while the second and third system qubits are in the ground state $|0\rangle$, i.e., $|1\rangle\otimes |0\rangle\otimes |0\rangle$, as illustrated schematically in Fig.~\ref{fig:5bits}(a).
We use this end-localized initial state because it allows the dissipation to induce a magnetization current from the first to the third qubit.

For reference, we numerically solve the Lindblad master equation and execute the same circuits on the noiseless AerSimulator. 
These reference results are shown as solid lines and open symbols in Figs.~\ref{fig:5bits}(b)--\ref{fig:5bits}(g), respectively.
The AerSimulator results exhibit small deviations from the Lindblad master-equation results owing to the finite time step $\Delta t$, shot noise, and the layered implementation of the three-qubit gates, but these discrepancies remain within an acceptable range.

For \emph{Reimei}, the raw experimental data already capture the qualitative trend of the magnetization current, although they deviate noticeably from the exact Lindblad dynamics, as shown in Fig.~\ref{fig:5bits}(b). 
After applying post-selection \ac{QEM}, the agreement is substantially improved, and the magnetization-current dynamics are accurately reproduced, as shown in Fig.~\ref{fig:5bits}(e). 
Although the number of effective shots is reduced from $N_{\mathrm{shots}} = 1024$ to approximately 600, the estimated shot noise remains small compared with the deviations from the master-equation results.

In contrast, on \emph{ibm\_kobe}, the measured dynamics deviate from the Lindblad master-equation results after only a few time steps, as shown in Fig.~\ref{fig:5bits}(c). 
Even after applying post-selection \ac{QEM}, discrepancies persist, as shown in Fig.~\ref{fig:5bits}(f). 
The number of shots is $N_{\mathrm{shots}}=4096$, which is four times larger than that used for \emph{Reimei}. 
Even after post-selection \ac{QEM}, the effective number of shots remains larger than that for \emph{Reimei}.

Finally, on \emph{ibm\_boston}, the measured dynamics show some improvement over those obtained on \emph{ibm\_kobe}, as shown in Fig.~\ref{fig:5bits}(d), but still exhibit noticeable deviations from the Lindblad master-equation results. 
Post-selection \ac{QEM} further improves the results, as shown in Fig.~\ref{fig:5bits}(g), although the overall accuracy remains below that achieved on \emph{Reimei}.

\subsection{Experimental results for $M=7$}

\begin{figure*}[htbp]
\includegraphics{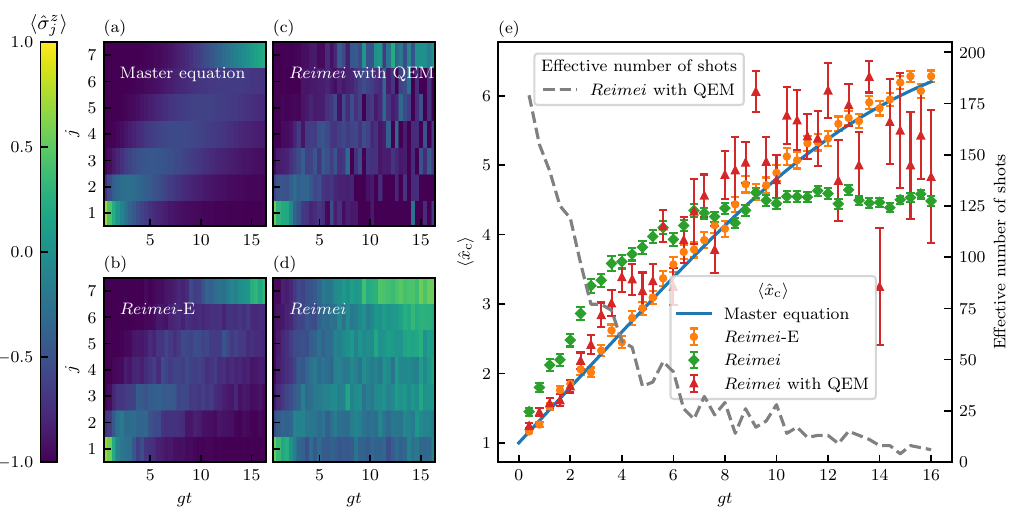}%
\caption{
    Results for the nonlocal dissipation with $M=7$ system qubits and time-step parameter $g\Delta t = 0.4$. 
    The number of shots is $N_{\mathrm{shots}} = 200$.
    Panels (a)--(d) show the magnetization-current dynamics of $\langle\hat{\sigma}^z_j\rangle$ obtained from the Lindblad master equation, 
    the noiseless \emph{Reimei}-E simulation, 
    \emph{Reimei} with post-selection \ac{QEM}, 
    and \emph{Reimei} without \ac{QEM}, respectively. 
    Panels (a)--(d) share the same axes ranges and color scale. 
    Panel (e) shows the time evolution of the center-of-mass position $\langle \hat{x}_{\mathrm{c}}\rangle$ extracted from the data in panels (a)--(d). 
    The symbols represent the corresponding results, and the error bars quantify shot noise as described in Appendix~\ref{sec:estimators}.
    The gray dashed line in panel (e), plotted against the right axis, indicates the effective number of shots remaining after post-selection \ac{QEM}.
    \label{fig:13bits}}
\end{figure*}

Since the $M = 3$ experiment on \emph{Reimei} accurately reproduces the solution of the Lindblad master equation, we next consider a larger case with $M=7$ on \emph{Reimei}, using six ancillas and 13 qubits in total. 
As in the $M=3$ case, the initial state is chosen such that the first system qubit is in the excited state $|1\rangle$ and all other system qubits are in the ground state $|0\rangle$, allowing us to observe a pronounced magnetization current. 
Because execution on the trapped-ion quantum computer \emph{Reimei} is costly, we first use its emulator, \emph{Reimei}-E, to run the same circuits noiselessly before hardware execution. 
This allows us to verify that the chosen time-step size $\Delta t$ and the layered implementation do not cause significant deviations. 
We increase the total simulated time to $gt=16$ to observe the extended dynamics of the magnetization flow. 
In addition, \emph{Reimei}-E provides an estimate of the runtime on the actual quantum device \emph{Reimei}. 
Based on these estimates, we reduce the number of shots to $N_{\mathrm{shots}}=200$ to shorten the hardware execution time. 
The noiseless \emph{Reimei}-E simulation results for these parameters are shown in Fig.~\ref{fig:13bits}(b) as a reference.

The results obtained from the Lindblad master equation, the \emph{Reimei}-E emulator, \emph{Reimei} with and without post-selection \ac{QEM} are shown in Figs.~\ref{fig:13bits}(a), \ref{fig:13bits}(b), \ref{fig:13bits}(c), and \ref{fig:13bits}(d), respectively.
The \emph{Reimei}-E emulator results show a pronounced magnetization-current signal from site 1 to site 7, 
and agree well with the Lindblad master-equation results, indicating that the quantum collision model remains accurate over $40$ time steps and that the layered implementation of the three-qubit gates remains reliable. 
On real hardware, the magnetization-current signal becomes weaker, as shown in Fig.~\ref{fig:13bits}(d), although the overall trend remains visible. 
Post-selection \ac{QEM} significantly improves the visibility of the magnetization current, as shown in Fig.~\ref{fig:13bits}(c).

To quantify the agreement between our experimental results and the numerical results obtained from the Lindblad master equation, we define the center-of-mass position. 
After mapping the qubit chain onto a fermionic lattice via the Jordan-Wigner transformation in Eq.~\eqref{eq:j-w-trans}, this quantity is given by 
\begin{align}\label{eq:com}
    \hat{x}_{\mathrm{c}} = \frac{\sum_{j=1}^{M}  \hat{n}_j j}
    {\sum_{j=1}^{M} \hat{n}_j}.
\end{align}
Note that $\hat{x}_{\mathrm{c}}$ becomes ill-defined when the denominator vanishes, i.e., when the total fermion number is zero. 
Therefore, even without \ac{QEM}, shots for which $\hat{x}_{\mathrm{c}}$ is ill-defined must be discarded. 
Details are provided in Appendix~\ref{sec:further-exp-data}.

The time evolution of the center-of-mass position $\hat{x}_{\mathrm{c}}$ is shown in Fig.~\ref{fig:13bits}(e). 
As a more compact quantitative measure, $\hat{x}_{\mathrm{c}}$ makes the effect of post-selection \ac{QEM} even more evident. 
In particular, the post-selected \emph{Reimei} results agree better with the Lindblad master-equation results than the raw results, until the effective shot count becomes limiting around $gt = 12$. 
This agreement persists even after the effective number of shots has fallen below 50.

\begin{figure}[htbp]
\includegraphics[width=\linewidth]{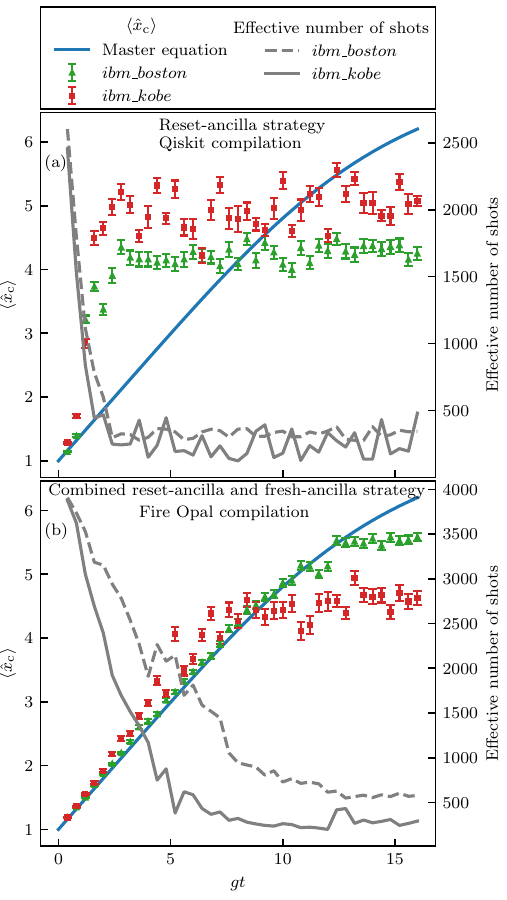}%
\caption{
    Results for the nonlocal dissipation model with $M=7$ system qubits obtained on \emph{ibm\_kobe} and \emph{ibm\_boston}. 
    Panel (a) shows results obtained using the reset-ancilla strategy with Qiskit compilation, while panel (b) shows results obtained using the combined reset-ancilla and fresh-ancilla strategy with Fire Opal compilation. 
    The red and green symbols show the center-of-mass position $\langle \hat{x}_{\mathrm{c}}\rangle$, with error bars quantifying shot noise as described in Appendix~\ref{sec:estimators}. 
    The gray dashed and solid lines indicate the effective number of shots after post-selection for \emph{ibm\_boston} and \emph{ibm\_kobe}, respectively. 
    A numerical solution of the Lindblad master equation is shown in blue for reference. 
    \label{fig:13bits-compare}}
\end{figure}

We then run the same circuits on \emph{ibm\_kobe} and \emph{ibm\_boston}. 
The corresponding dynamics of the center-of-mass position $\hat{x}_{\mathrm{c}}$ is shown in Fig.~\ref{fig:13bits-compare}(a).
The system information is lost rapidly after only a few time steps, even with \ac{QEM}. 
At the same time, the effective number of shots decreases from the initial value of $N_{\mathrm{shots}} = 4096$ to below 500 after only a few steps. 
This suggests that, on current superconducting quantum hardware, such deep circuits with many reset operations become dominated by noise, because the circuit duration quickly becomes comparable to or exceeds the coherence time $T_1$ of the superconducting qubits.

To match the performance of the \emph{Reimei} device shown in Fig.~\ref{fig:13bits}, it was necessary to drastically reduce the circuit duration. 
This required reducing both the number of reset operations and the gate depth of the circuits. 
The large number of  qubits available on the Heron processors enables the use of different ancilla strategies. 
We therefore applied a hybrid strategy combining fresh and reset ancillas to balance the number of required reset operations against the number of ancillas. 
For $gt\in(0,12]$, we reset all ancillas every fifteenth layer, whereas for $gt\in(12, 16]$, we reset them every twentieth layer. 
This strategy uses up to $20 \times 6 = 120$ ancillas.

To reduce the circuit duration, we use the compiler in Q-CTRL's Fire Opal software~\cite{fire_opal}. This software reduces the required number of quantum gates through optimized logical transpilation and identifies suitable qubit layouts for the circuits considered here, thereby reducing the need for additional SWAP gates during routing. 
Details of the compiled circuit properties are provided in Appendix~\ref{sec:circuit-detail}.

The gate count can be further reduced by removing three-qubit unitaries \(U_3(j)\) that act only on qubits remaining in the ground state. 
Such unitaries act trivially on the corresponding input state and therefore do not affect the dynamics. 
In our initial state, only the first system qubit is excited, while all other system qubits and ancillas are in the ground state. 
Since the excitation propagates to the right by at most two sites per time step of the quantum \ac{ASEP} circuit, many three-qubit gates acting far from the excitation front are redundant in the early time steps. 
We therefore remove these redundant $U_3(j)$ gates in the first three time steps of the circuit, reducing the total gate count without changing the ideal noiseless dynamics.

The results obtained with the optimized circuits are shown in Fig.~\ref{fig:13bits-compare}(b). 
The measured center-of-mass positions now closely follow the Lindblad master-equation results over a wide range of simulated durations $gt$, and the effective number of shots decreases much more slowly. 
The results obtained on \emph{ibm\_boston} are further improved compared with those obtained on \emph{ibm\_kobe}, reflecting the hardware improvement from the second to third generation of IBM Heron processors.

With optimized circuit engineering and compilation, the platform-specific challenges of superconducting quantum processors can be mitigated while their strengths are exploited. 
The fast gate times of superconducting processors allow the simulations to be executed with a large number of shots. 
In combination with error suppression provided by Fire Opal, which increases the fraction of effective shots after \ac{QEM}, this provides sufficient statistics to interpret the experimental results. 
In addition, the ability to execute many operations in parallel enables compilation to shallower circuits.

\section{Summary and Outlook}

In this work, we presented a framework for simulating quantum collision models on quantum computers, focusing on two ancilla strategies. 
The reset-ancilla strategy reuses a small number of ancillas at the cost of frequent reset operations, which can lead to long circuit durations and additional reset-induced noise, including crosstalk, on some hardware platforms. 
The fresh-ancilla strategy avoids reset operations but requires a larger number of ancillas, which grows in proportion to the number of simulated time steps. 
These two strategies can also be combined. The optimal choice of strategy depends on the hardware characteristics of the quantum computer, with the accuracy of reset operations and the number of available qubits being the most important factors.

We tested the limitations of currently available \ac{NISQ} hardware using one trapped-ion quantum computer, \emph{Reimei}, and two superconducting quantum computers, \emph{ibm\_kobe} and \emph{ibm\_boston}. 
We implemented collision models for three physical systems of increasing size, ranging from one to seven system qubits. 
As reference results, we numerically solved the Lindblad master equations corresponding to the three models. 
To quantify approximations introduced in the circuit construction, such as those arising from Trotterization and the layered implementation of three-qubit gates, we also executed all circuits on noiseless simulators.

The first experiment modeled local dissipation in a two-level system coupled to a bosonic field. 
Our hardware results reproduced the predicted spontaneous emission and coherent decay. 
This model was implemented using two qubits and the reset-ancilla strategy. 
The results obtained on \emph{Reimei} and \emph{ibm\_boston} showed only small deviations from the exact solution, whereas the experiment on \emph{ibm\_kobe} was more strongly affected by noise associated with reset operations.

The second experiment simulated a quantum \ac{ASEP} with three system qubits and two ancillas, using the reset-ancilla strategy for all circuits. 
We developed a post-selection \ac{QEM} method based on magnetization conservation to mitigate the effect of bit flip errors, which substantially improved the results in all cases. 
We observed strong noise effects on both superconducting quantum computers, with \emph{ibm\_boston} again outperforming \emph{ibm\_kobe}, whereas \emph{Reimei} closely reproduced the theoretical results after \ac{QEM} was applied.

The third experiment simulated a larger quantum \ac{ASEP} with seven system qubits for up to 40 time steps. 
The results obtained on \emph{Reimei} still captured the magnetization current after post-selection \ac{QEM}, although shot noise led to large uncertainties because of the limited number of experimental repetitions and the low fraction of valid measurements after \ac{QEM}. 
On the superconducting quantum computers, the results became dominated by noise after only a few simulated time steps when the reset-ancilla strategy and default compilation were used. 
We therefore employed an optimized combination of the reset-ancilla and fresh-ancilla strategies to reduce the noise associated with reset operations, together with the compiler in Q-CTRL's Fire Opal software to reduce the two-qubit gate depth of the circuit by about an order of magnitude; see Appendix~\ref{sec:circuit-detail}. 
With these optimized circuits, \emph{ibm\_kobe} showed improved results, while \emph{ibm\_boston} achieved the best agreement with the exact solution, even for larger numbers of time steps.

The presented results indicate that the performance of \ac{NISQ} devices can be significantly improved by designing hardware-specific strategies and using optimized compilers. 
We demonstrated that the strengths of superconducting quantum computers, namely their large qubit counts and ability to execute many two-qubit gates in parallel, can be exploited to mitigate the challenge posed by noisy reset operations.

In recent years, quantum computing hardware has advanced remarkably rapidly. 
In particular, the realization of a 98-qubit trapped-ion quantum computer~\cite{ransford2025helios98qubittrappedionquantum} opens the possibility of simulating larger and more intricate open quantum systems. 
The all-to-all connectivity available in trapped-ion platforms may also be advantageous for implementing more complex dissipative interactions, including those arising in higher-dimensional models. 
On the software side, advances in circuit compilation and optimization~\cite{qctrl2026practicaladvantage} are expected to be transferrable to open-system simulations and may further expand the range of dissipative processes that can be simulated. 
These hardware and software developments make quantum collision models a promising framework for studying open quantum systems in regimes that are challenging for classical simulation.

%



\begin{acknowledgments}
We acknowledge the use of QuTiP~\cite{qutip5}, a convenient python package, for numerically solving the Lindblad master equation. 
This work is partially based on results obtained from Project No.~JPNP20017, supported by the New Energy and Industrial Technology Development Organization (NEDO). 
This study was also supported by MEXT, Japan, through JSPS KAKENHI Grant No.~JP21H04446, JST COI-NEXT Grant No.~JPMJPF2221, and the Program for Promoting Research on the Supercomputer Fugaku under Grant No.~MXP1020230411. 
We also acknowledge support from the UTokyo Quantum Initiative, the RIKEN TRIP initiative (RIKEN Quantum), and the COE research grant in computational science from Hyogo Prefecture and Kobe City through the Foundation for Computational Science. 
M. O. T. acknowledges support from the JSPS Research Fellowship for Young Scientists.

\end{acknowledgments}

\appendix

\section{Derivation of the Lindblad master equation from quantum collision models\label{sec:proof}}

Here we derive the Lindblad master equation from the quantum collision model for the Lindblad form used in this work, in which each jump operator $\hat{L}_j$ is associated with an independent dissipation rate $\Gamma_j$. 
We choose the characteristic energy scale of the system Hamiltonian, denoted by $\mathcal{O}(\hat{H})$, as the energy unit, so that $\mathcal{O}(\hat{H})\sim \mathcal{O}(1)$. 
We also assume that the dissipation rates $\Gamma_j$ remain finite in the continuous-time limit $\Delta t \to 0$. 
From Eq.~\eqref{eq:conditions}, the coupling strength then scales as $g_j = \sqrt{\Gamma_j}\Delta t^{-1/2}$. 
Thus, in the limit $\Delta t\to 0$, the time step $\Delta t$ is a small parameter, while $g_j$ is a large parameter of order $\mathcal{O}(\Delta t^{-1/2})$. 
Accordingly, the expansion of the collision map must be organized carefully in powers of $\sqrt{\Delta t}$.

Using $g_j^2\Delta t=\Gamma_j$, we rewrite the system-ancilla interaction as 
\begin{align}
\hat{V}
= \frac{\hat{v}}{\sqrt{\Delta t}},
\quad
\hat{v}
= \sum_j \sqrt{\Gamma_j}
\left(
\hat{L}_j\hat{\sigma}^{+}_{\mathrm{a},j}
+ \hat{L}_j^{\dagger}\hat{\sigma}^{-}_{\mathrm{a},j}
\right).
\end{align}
The operator $\hat{v}$ remains finite in the continuous-time limit $\Delta t\to 0$. 
The evolution operator for a single time step is then
\begin{align}
U(\Delta t)
=
e^{-i\hat{H}\Delta t}
e^{-i\hat{v}\sqrt{\Delta t}}.
\end{align}
Expanding this single-step unitary up to order $\Delta t$, we obtain
\begin{align}\nonumber
U(\Delta t)
=&
\left[
1-i\hat{H}\Delta t
+\mathcal{O}(\Delta t^2)
\right] \\ \nonumber
& \times\left[
1-i\hat{v}\sqrt{\Delta t}
-\frac{1}{2}\hat{v}^2\Delta t
+\mathcal{O}(\Delta t^{3/2})
\right] \\ 
=&
1-i\hat{v}\sqrt{\Delta t}
-i\hat{H}\Delta t
-\frac{1}{2}\hat{v}^2\Delta t
+\mathcal{O}(\Delta t^{3/2}).
\end{align}
Here, the cross term between $\hat{H}\Delta t$ and $\hat{v}\sqrt{\Delta t}$ is of order $\Delta t^{3/2}$ and is therefore included in the remainder.

The quantum collision map is defined by
\begin{align}
\mathcal{E}[\hat{\rho}(t)]
=
\mathrm{Tr}_{\mathrm a}
\left[
U(\Delta t)
\left(
\hat{\rho}(t)\otimes\hat{\rho}_{\mathrm a}
\right)
U^{\dagger}(\Delta t)
\right],
\end{align}
where $\hat{\rho}_{\mathrm a}$ is the state to which the ancillas are reset after each collision. 
Substituting the expansion of $U(\Delta t)$ and keeping terms up to order $\Delta t$, we find
\begin{align}\nonumber
\mathcal{E}[\hat{\rho}(t)]
=&
\hat{\rho}(t)
-i\sqrt{\Delta t}\,
\mathrm{Tr}_{\mathrm a}
\left[
\hat{v},
\hat{\rho}(t)\otimes\hat{\rho}_{\mathrm a}
\right] 
-i\Delta t
[\hat{H},\hat{\rho}(t)]
\\ \nonumber
&
-\frac{\Delta t}{2}
\mathrm{Tr}_{\mathrm a}
\left\{
\hat{v}^2,
\hat{\rho}(t)\otimes\hat{\rho}_{\mathrm a}
\right\}
+\Delta t\,
\mathrm{Tr}_{\mathrm a}
\left[
\hat{v}
\left(
\hat{\rho}(t)\otimes\hat{\rho}_{\mathrm a}
\right)
\hat{v}
\right]
\\ \label{eq:map-series}
&
+\mathcal{O}(\Delta t^{3/2}).
\end{align}
Here, the term proportional to $\sqrt{\Delta t}$ comes from the first-order contribution in $\hat{v}$, while the two terms proportional to $\Delta t$ involving $\hat{v}^2$ arise from the second-order expansion of the system-ancilla interaction.

In this work, each ancilla is reset to the ground state,
\begin{align}
\hat{\rho}_{\mathrm a}
=
\bigotimes_j
|0_{\mathrm a,j}\rangle
\langle 0_{\mathrm a,j}|.
\end{align}
This state has no coherence between $|0_{\mathrm a,j}\rangle$ and $|1_{\mathrm a,j}\rangle$.
As a result, all terms containing an odd number of $\hat{v}$ operators vanish after tracing over the ancillas.
In particular,
\begin{align}
\mathrm{Tr}_{\mathrm a}
\left[
\hat{v},
\hat{\rho}(t)\otimes\hat{\rho}_{\mathrm a}
\right]
=0.
\end{align}
The remaining second-order terms can be evaluated using the ancilla trace  identities 
\begin{align}
\mathrm{Tr}_{\mathrm a}
\left[
\hat{\sigma}^{-}_{\mathrm a,j}
\hat{\sigma}^{+}_{\mathrm a,k}
\hat{\rho}_{\mathrm a}
\right]
=\delta_{jk},
\quad
\mathrm{Tr}_{\mathrm a}
\left[
\hat{\sigma}^{+}_{\mathrm a,j}
\hat{\sigma}^{-}_{\mathrm a,k}
\hat{\rho}_{\mathrm a}
\right]
=0.
\end{align}
Thus, cross terms with different ancilla indices vanish because a nonzero contribution must excite and de-excite the same ancilla.
Therefore,
\begin{align}
\mathrm{Tr}_{\mathrm a}
\left\{
\hat{v}^2,
\hat{\rho}(t)\otimes\hat{\rho}_{\mathrm a}
\right\}
=
\sum_j
\Gamma_j
\left\{
\hat{L}_j^{\dagger}\hat{L}_j,
\hat{\rho}(t)
\right\},
\end{align}
and
\begin{align}
\mathrm{Tr}_{\mathrm a}
\left[
\hat{v}
\left(
\hat{\rho}(t)\otimes\hat{\rho}_{\mathrm a}
\right)
\hat{v}
\right]
=
\sum_j
\Gamma_j
\hat{L}_j
\hat{\rho}(t)
\hat{L}_j^{\dagger}.
\end{align}
Moreover, since all half-integer orders in $\Delta t$ contain an odd number of $\hat{v}$ operators, they also vanish after the partial trace over the ancillas.
Thus, the remainder $\mathcal{O}(\Delta t^{3/2})$ in Eq.~\eqref{eq:map-series} can be replaced by $\mathcal{O}(\Delta t^2)$ for the present ancilla state.

Substituting these results into Eq.~\eqref{eq:map-series}, we obtain
\begin{align}\nonumber
\mathcal{E}[\hat{\rho}(t)]
=&
\hat{\rho}(t)
-i[\hat{H},\hat{\rho}(t)]\Delta t
\\ \nonumber
&+\sum_j \Gamma_j
\left[
\hat{L}_j\hat{\rho}(t)\hat{L}_j^{\dagger}
-\frac{1}{2}
\left\{
\hat{L}_j^{\dagger}\hat{L}_j,
\hat{\rho}(t)
\right\}
\right]\Delta t \\ \nonumber
&+\mathcal{O}(\Delta t^2)
\\
=&
\hat{\rho}(t)
+\mathcal{L}[\hat{\rho}(t)]\Delta t
+\mathcal{O}(\Delta t^2).
\end{align}
It follows that
\begin{align}
\frac{
\mathcal{E}[\hat{\rho}(t)]-\hat{\rho}(t)
}{\Delta t}
=
\mathcal{L}[\hat{\rho}(t)]
+\mathcal{O}(\Delta t).
\end{align}
Taking the continuum limit $\Delta t\to 0$, the quantum collision model therefore yields the Lindblad master equation,
\begin{align}
\frac{\mathrm d}{\mathrm dt}\hat{\rho}(t)
=
\lim_{\Delta t\to 0}
\frac{
\mathcal{E}[\hat{\rho}(t)]-\hat{\rho}(t)
}{\Delta t}
=
\mathcal{L}[\hat{\rho}(t)].
\end{align}

\section{Statistical estimators and error bars\label{sec:estimators}}

In this section, we summarize the statistical estimators used in this work and explain how the shot-noise error bars are evaluated.

Let the system be in a state $\hat{\rho}$, and let $\hat{O}$ be an observable whose expectation value is
\begin{align}
\langle \hat{O} \rangle = \mathrm{Tr}(\hat{O}\hat{\rho}).
\end{align}
Suppose that $\hat{O}$ has the spectral decomposition
\begin{align}
\hat{O} = \sum_m O_m |m\rangle\langle m|,
\end{align}
where $O_m$ is the eigenvalue associated with the eigenstate $|m\rangle$.
A projective measurement of $\hat{O}$ gives the outcome $O_m$ with probability
\begin{align}
p_m = \mathrm{Tr}\left[
|m\rangle\langle m| \hat{\rho}
\right].
\end{align}
Equivalently, a single measurement outcome can be regarded as a random variable $X_{\hat{O}}$ that takes the value $O_m$ with probability $p_m$.
Its expectation value is
\begin{align}
E[X_{\hat{O}}]
= \sum_m O_m p_m
= \mathrm{Tr}(\hat{O}\hat{\rho})
= \langle \hat{O} \rangle,
\end{align}
and its variance is
\begin{align}
\mathrm{var}[X_{\hat{O}}]
=
E[X_{\hat{O}}^2] - E[X_{\hat{O}}]^2
=
\langle \hat{O}^2\rangle - \langle \hat{O}\rangle^2.
\end{align}

In an experiment, the same circuit is repeated $N_{\mathrm{shots}}$ times.
Let $X_{\hat{O},l}$ denote the outcome of the $l$th shot.
Assuming independent and identically distributed shots, the sample-mean estimator is
\begin{align}
\bar{X}_{\hat{O}}
=
\frac{1}{N_{\mathrm{shots}}}
\sum_{l=1}^{N_{\mathrm{shots}}}
X_{\hat{O},l}.
\end{align}
This estimator is unbiased for the expectation value of the actually measured state:
\begin{align}
E[\bar{X}_{\hat{O}}]
=
\langle \hat{O}\rangle.
\end{align}
Because different shots are assumed to be independent, the variance of the sum is the sum of the variances.
Moreover, since all shots are generated by the same circuit, the random variables $X_{\hat{O},l}$ are identically distributed. 
Therefore, the variance of the sample-mean estimator is 
\begin{align}
\mathrm{var}[\bar{X}_{\hat{O}}]
=
\frac{1}{N_{\mathrm{shots}}}
\mathrm{var}[X_{\hat{O}}].
\end{align}
Thus, the statistical uncertainty due to finite sampling decreases as $1/\sqrt{N_{\mathrm{shots}}}$.

On noisy quantum hardware, the actually measured state $\hat{\rho}$ is generally different from the ideal noiseless state $\hat{\rho}_{\mathrm{ideal}}$.
The estimator may therefore be biased with respect to the ideal expectation value
\begin{align}
\langle \hat{O} \rangle_{\mathrm{ideal}}
=
\mathrm{Tr}(\hat{O}\hat{\rho}_{\mathrm{ideal}}).
\end{align}
For the sample-mean estimator, this bias is
\begin{align}
\mathrm{bias}[\bar{X}_{\hat{O}}]
=
E[\bar{X}_{\hat{O}}]
-
\langle \hat{O} \rangle_{\mathrm{ideal}}.
\end{align}
The mean-squared error is then
\begin{align}
\mathrm{MSE}[\bar{X}_{\hat{O}}]
=&
E\left[
\left(
\bar{X}_{\hat{O}}
-
\langle \hat{O} \rangle_{\mathrm{ideal}}
\right)^2
\right] \nonumber \\
=&
\mathrm{bias}[\bar{X}_{\hat{O}}]^2
+
\mathrm{var}[\bar{X}_{\hat{O}}].
\end{align}
Quantum error mitigation aims to reduce the bias, typically at the cost of increasing the variance.

We now specialize to the Pauli observable used throughout the experiments.
With our convention,
\begin{align}
\hat{\sigma}^z
=
|1\rangle\langle 1|
-
|0\rangle\langle 0|,
\end{align}
the eigenvalue is $-1$ for the outcome $|0\rangle$ and $+1$ for the outcome $|1\rangle$.
Thus, a single-shot measurement of $\hat{\sigma}^z$ is described by a random variable $X_{\hat{\sigma}^z}$ satisfying
\begin{align}
X_{\hat{\sigma}^z}(|0\rangle)=-1,
\quad
X_{\hat{\sigma}^z}(|1\rangle)=+1.
\end{align}
The corresponding probabilities are
\begin{align}
P(X_{\hat{\sigma}^z}=-1)
=
\langle 0|\hat{\rho}|0\rangle,
\quad
P(X_{\hat{\sigma}^z}=+1)
=
\langle 1|\hat{\rho}|1\rangle.
\end{align}
Therefore,
\begin{align}
    E[X_{\hat{\sigma}^z}] =& P(X_{\hat{\sigma}^z} = 1)
    - P(X_{\hat{\sigma}^z} = -1) \nonumber \\
    =& \mathrm{Tr}\left[
        \hat{\sigma}^z \hat{\rho}
    \right] = \langle \hat{\sigma}^z \rangle    .
\end{align}
and, since $(\hat{\sigma}^z)^2=1$,
\begin{align}
\mathrm{var}[X_{\hat{\sigma}^z}]
=
1-\langle \hat{\sigma}^z\rangle^2.
\end{align}
For $N_{\mathrm{shots}}$ shots, the sample-mean estimator
\begin{align}
\bar{X}_{\hat{\sigma}^z}
=
\frac{1}{N_{\mathrm{shots}}}
\sum_{l=1}^{N_{\mathrm{shots}}}
X_{\hat{\sigma}^z,l}
\end{align}
has variance
\begin{align}
\mathrm{var}[\bar{X}_{\hat{\sigma}^z}]
=
\frac{
1-\langle \hat{\sigma}^z\rangle^2
}{
N_{\mathrm{shots}}
}.
\end{align}
In practice, we estimate this variance by replacing $\langle \hat{\sigma}^z\rangle$ with the measured sample mean $\bar{X}_{\hat{\sigma}^z}$:
\begin{align}
\widehat{\mathrm{SE}}(\bar{X}_{\hat{\sigma}^z})
=
\sqrt{
\frac{
1-\bar{X}_{\hat{\sigma}^z}^{2}
}{
N_{\mathrm{shots}}
}},
\end{align}
where $\widehat{\mathrm{SE}}$ denotes the estimated standard error.
These standard errors are shown as the shot-noise error bars for the $\langle \hat{\sigma}^z_j\rangle$ data.

For the post-selection \ac{QEM} used in the quantum \ac{ASEP} experiments, only shots satisfying the conserved-value condition for the total $z$-magnetization are retained.
Let $N_{\mathrm{eff}}$ be the number of shots remaining after post-selection.
The same estimator is then applied to the retained shots, with $N_{\mathrm{shots}}$ replaced by $N_{\mathrm{eff}}$.
Thus, for post-selected $\hat{\sigma}^z$ measurements, we use
\begin{align}
\widehat{\mathrm{SE}}_{\mathrm{post}}(\bar{X}_{\hat{\sigma}^z})
=
\sqrt{
\frac{
1-\bar{X}_{\hat{\sigma}^z}^{2}
}{
N_{\mathrm{eff}}
}}.
\end{align}
This increase in the standard error reflects the reduced number of effective shots after post-selection.

For the center-of-mass position $\hat{x}_{\mathrm c}$ defined in Eq.~\eqref{eq:com}, the estimator is computed directly from the measured bitstrings.
For each retained shot $l$, we compute a value $x_{\mathrm c,l}$ from the corresponding occupation configuration. 
Shots for which $\hat{x}_{\mathrm c}$ is ill-defined, namely those with zero total particle number, are discarded as described in Appendix~\ref{sec:further-exp-data}. 
Let $N_{\mathrm{eff}}$ denote the number of shots retained after this filtering and, when applicable, post-selection \ac{QEM}. 
The center-of-mass estimator is the sample mean
\begin{align}
\bar{x}_{\mathrm c}
=
\frac{1}{N_{\mathrm{eff}}}
\sum_{l=1}^{N_{\mathrm{eff}}}
x_{\mathrm c,l}, 
\end{align}
and the corresponding sample variance is
\begin{align}
s^2_{x_{\mathrm c}} = 
\frac{1}{N_{\mathrm{eff}}}
\sum_{l=1}^{N_{\mathrm{eff}}}
\left(
x_{\mathrm c,l} - 
\bar{x}_{\mathrm c}
\right)^2.
\end{align}
The shot-noise error bar is then estimated as the standard error of the sample mean,
\begin{align}
\widehat{\mathrm{SE}}(\bar{x}_{\mathrm c}) = 
\sqrt{
\frac{s^2_{x_{\mathrm c}}}{N_{\mathrm{eff}}}
} .
\end{align}
This expression is used for the error bars of the center-of-mass data shown in this work.

\section{Hybrid ancilla strategies for quantum \ac{ASEP} dynamics \label{sec:hybrid_ancilla}}

\begin{figure*}[htbp]
    \centering
    \includegraphics[width=\linewidth]{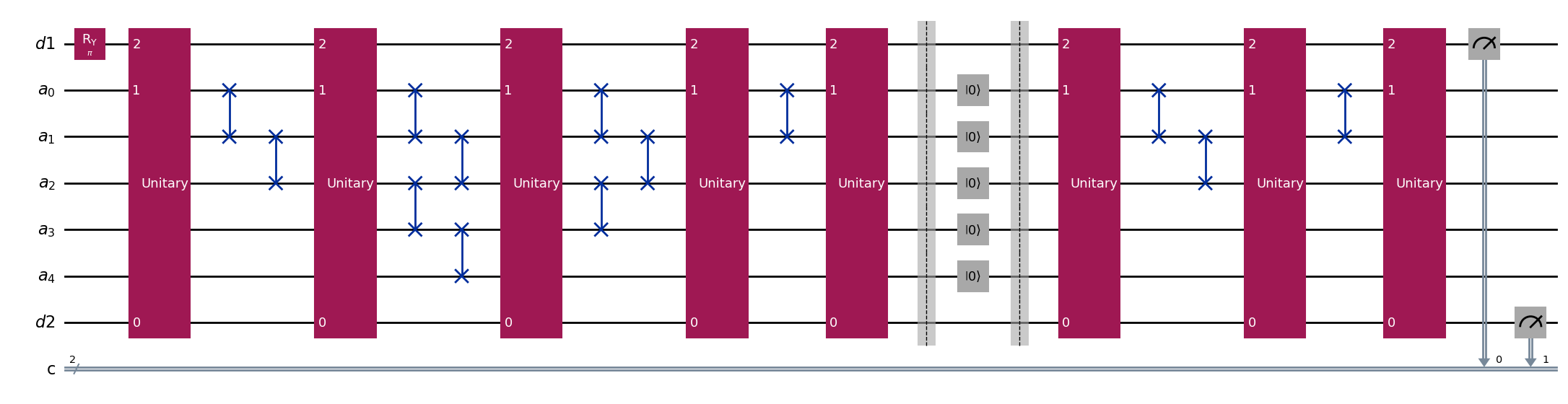}
    \caption{
    Example circuit illustrating a hybrid strategy that combines fresh ancillas with ancilla resets. 
    The circuit uses two system qubits, $d1$ and $d2$, and five ancillas, $a_0,\dots,a_4$, to implement repeated three-qubit interactions. 
    The labels $0$, $1$, and $2$ inside each unitary block indicate the three qubits on which the corresponding three-qubit unitary acts. 
    The ancillas are routed along a linear chain by SWAP operations arranged in a brick-wall pattern, so that a fresh ancilla is supplied to each system-ancilla interaction. 
    SWAP operations between two ancillas that are either both fresh or both already used are omitted. 
    After all five ancillas have been used, they are reset and reused for subsequent interactions. 
     }
    \label{fig:SWAP-chain}
\end{figure*}

The ancilla strategies shown schematically in Fig.~\ref{fig:collision-model} can be combined to optimize circuits for the IBM Heron processors. 
Fig.~\ref{fig:SWAP-chain} shows an example of such a hybrid strategy,  combining the fresh-ancilla and reset-ancilla strategies, for a circuit containing two system qubits and five ancillas. 
The circuit contains eight repeated three-qubit interactions, each of which requires an ancilla initialized in the ground state. 
The five ancillas are first used as fresh ancillas. 
After all five ancillas have participated in one interaction, 
they are reset and then routed again to the system qubits using the same pattern of SWAP gates. 
In this way, the number of available ancillas and the number of interaction layers determine how many reset operations are required. 
In the example shown in Fig.~\ref{fig:SWAP-chain}, two of the five reset operations are not strictly necessary and are eliminated by the compiler in Q-CTRL's Fire Opal software.

We control the routing of the ancillas by swapping them along a linear chain. The SWAP operations are arranged in a brick-wall pattern, which implements a cyclic permutation of the ancilla positions. 
SWAP operations between two ancillas that are either both still fresh or both already used but not yet reset are omitted, because exchanging such ancillas does not affect the subsequent interactions. 
Furthermore, each required SWAP operation is implemented using two CNOT gates, because one of the two qubits involved is known to be in the ground state.

The precise combination of fresh-ancilla usage and reset operations must be chosen carefully. 
Increasing the number of ancillas reduces the number of required reset operations, but it also increases the number of two-qubit gates needed for routing. 
On the Heron processors, one reset operation took roughly as long as 30 two-qubit gates, whereas swapping in a fresh ancilla increases the two-qubit gate depth by 2. 
For the data shown in Fig.~\ref{fig:13bits-compare}(b), we reset the ancillas every 15 layers for data points with $gt\leq 12$ and every 20 layers for $gt>12$. 
Thus, we used $15 \times 6 = 90$ ancillas for experiments with up to 30 layers and increased this number to $20 \times 6 = 120$ for $gt>12$. 
With this choice of ancilla strategy, no ancilla was reset more than once. 
The performance could be further optimized by systematically benchmarking different combinations of the number of ancillas and reset intervals.

\section{Circuit compilation details \label{sec:circuit-detail}}

\begin{figure}[htbp]
    \centering
    \includegraphics[width=\linewidth]{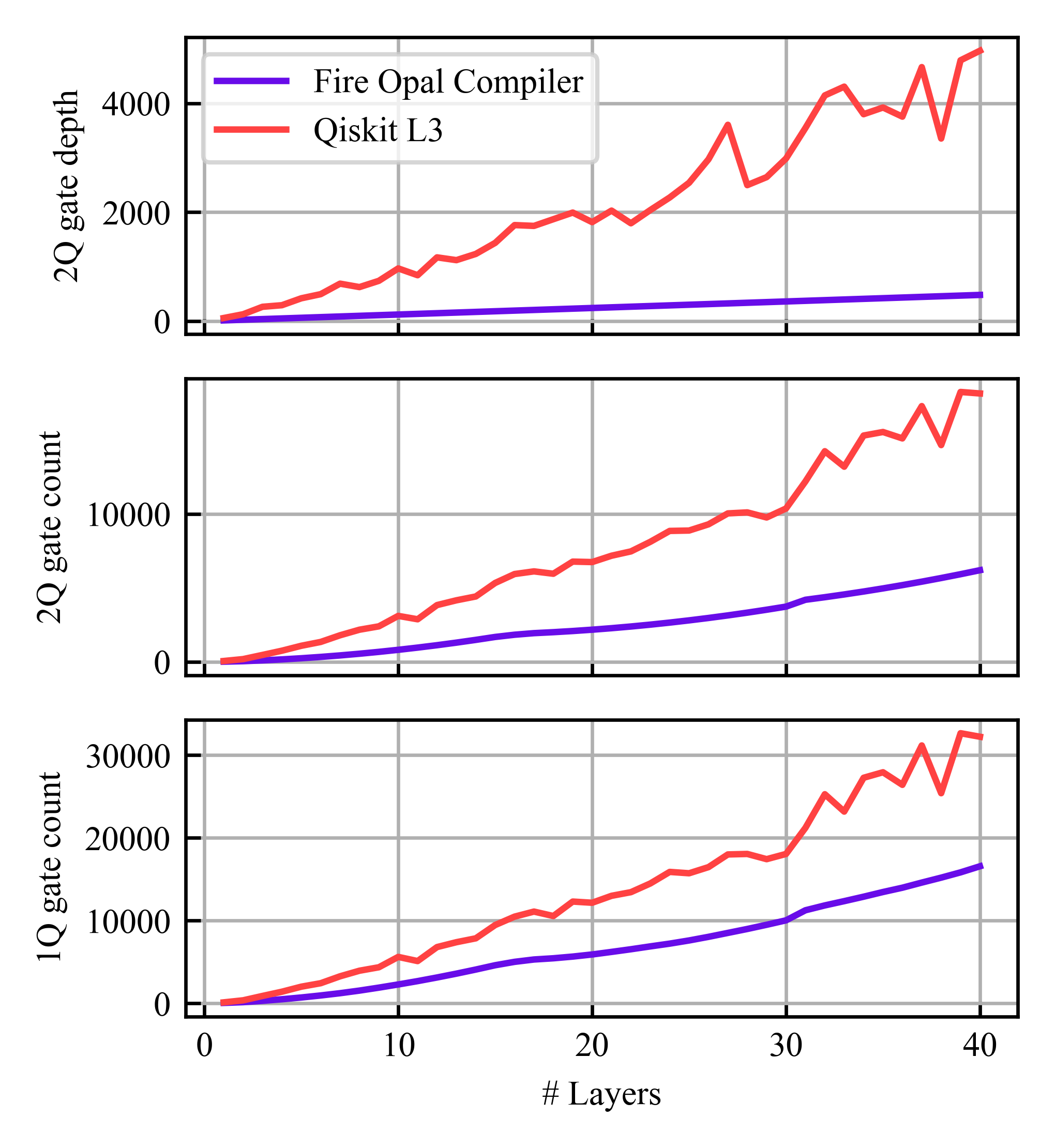}
    \caption{
    Analysis of compiled circuits used to obtain the experimental data shown in Fig.~\ref{fig:13bits-compare}(b). 
    The single-qubit gate counts include only physical single-qubit gates; virtual single-qubit gates, such as frame updates, are excluded because they do not appreciably affect the circuit duration. 
    The Q-CTRL Fire Opal compiler~\cite{fire_opal} substantially reduces both the two-qubit gate depth and the gate counts, making the experiments feasible on \ac{NISQ} hardware. 
     }
    \label{fig:superconducting_circ_props}
\end{figure}

Q-CTRL's Fire Opal software~\cite{fire_opal} was critical for achieving the performance shown in Fig.~\ref{fig:13bits-compare}(b). 
To investigate the influence of the transpiler, we compare the number of operations and the two-qubit gate depth with those of circuits transpiled using Qiskit with optimization level 3, as shown in Fig.~\ref{fig:superconducting_circ_props}. 
We find that the compiler in Q-CTRL's Fire Opal software produces much shallower circuits with fewer gates. 
The two most important factors for the circuit performance are the synthesis of the three-qubit unitaries and the layout selection. 
The logical compilation substantially reduces the gate depth through optimized synthesis of the three-qubit unitaries. 
In the routing step, the Fire Opal layout-selection functionality identifies a layout for which no additional routing is necessary, because the required connectivity is isomorphic to a subgraph of the device topology.

All circuits for experiments on the \emph{Reimei} quantum computer were compiled using Quantinuum Nexus. The compiled circuits contain 114 two-qubit gates and 174 physical single-qubit gates per layer; virtual single-qubit gates, such as frame updates, are not included in this count because they do not appreciably affect the circuit duration. 
Dense circuits such as those arising from this collision model necessarily become deeper after compilation for \emph{Reimei}, because the H1 processors can execute at most five two-qubit gates simultaneously.

\section{Additional experimental data \label{sec:further-exp-data}}

\begin{figure}
    \centering
    \includegraphics[width=\linewidth]{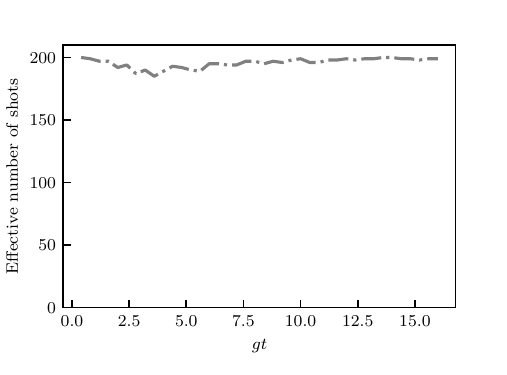}
    \caption{
    Effective number of shots for the \emph{Reimei} data without post-selection \ac{QEM} shown in Fig.~\ref{fig:13bits}(e), after discarding outcomes with zero total particle number. 
    }
    \label{fig:13-bits-effective-shots-noEM}
\end{figure}

Figure~\ref{fig:13-bits-effective-shots-noEM} shows the effective number of shots corresponding to the \emph{Reimei} data without post-selection \ac{QEM} in Fig.~\ref{fig:13bits}(e). 
Here, the effective number of shots refers to the number of measurement outcomes remaining after discarding shots for which the total particle number $\sum_j \hat{n}_j$ vanishes. 
These shots are discarded because the center-of-mass position defined in Eq.~\eqref{eq:com} is ill-defined when the total particle number is zero.

\section{List of acronyms\label{sec:acronyms}}

\begin{acronym}
    \acro{QEM}{quantum error mitigation}
    \acro{QES}{quantum error suppression}
    \acro{FTQC}{fault-tolerant quantum computer}
    \acro{NISQ}{noisy intermediate-scale quantum}
    \acro{ASEP}{asymmetric simple exclusion process}
    \acro{GKSL}{Gorini-Kossakowski-Sudarshan-Lindblad}
\end{acronym}

\bibliography{quantum-collision-model-on-QPU}

\end{document}